\documentclass{article}
\usepackage{graphicx}	% Including figure files
\usepackage{amsmath}	% Advanced maths commands

\usepackage{natbib}
\title{Meteor shower activity profiles and the use of orbital dissimilarity ($D$) criteria}
\author{Althea V. Moorhead\\NASA Meteoroid Environment Office,\\
Marshall Space Flight Center, Huntsville, AL 35812, USA}
\date{October 2019}

\begin{document}

\maketitle

\begin{abstract}
Orbital dissimilarity, or $D$, criteria are often used to select members of a meteor shower from a set of meteor observations. These criteria provide a quantitative description of the degree to which two orbits differ; if the degree of dissimilarity between a shower's reference orbit and an individual meteor does not exceed a selected threshold, the meteor is considered to be a member of that shower. However, members of a meteor shower tend to disperse in longitude of the ascending node (and thus in solar longitude) while preserving a common Sun-centered ecliptic radiant. Employing dissimilarity criteria to judge shower membership may therefore make the shower appear briefer than it actually is. We demonstrate this effect for two simulated meteor showers and assess the maximum permitted deviation in solar longitude as a function of radiant and velocity measurement error.
\end{abstract}

\section{Introduction}

Meteor showers are transient but typically annually recurring phenomena in which groups of meteors with similar orbits intersect the Earth. Because they have similar orbits, they tend to intersect the Earth with the same speed, radiant (or directionality), and solar longitude (time of year). The activity level of a shower is often described in terms of zenithal hourly rate, or ZHR, which is the number of visual meteors that would be seen under ideal observing conditions when the radiant is directly overhead. Thus, the basic set of parameters describing a meteor shower typically include a peak solar longitude, radiant, geocentric speed, and peak or maximum ZHR.

The peak or maximum ZHR describes the intensity of a meteor shower only during its most active night (or hour). The shower will also produce off-peak activity, albeit at a lower rate. Unless a shower exhibits multiple peaks, the ZHR increases steadily before and decays after the peak date and time. The ``sharpness'' of the peak, or the rise and decay time, can vary substantially between showers. For instance, the Quadrantids are a very brief shower, with measurable activity lasting only 2-3 days. In contrast, the Northern and Southern Taurids can last weeks, if not months. In most cases, the rise, peak, and decay in a shower's activity can be characterized as a double-exponential profile \citep{1990acm..proc..535J,1994A&A...287..990J,1999A&AS..135..291O,2002JIMO...30..168D,Moorhead2017}. 

An accurate description of the change in a meteor shower's activity over time has multiple uses. It can be used to plan observations, or to calibrate off-peak flux measurements. It is critical for generating meteor shower forecasts \citep{1997AdSpR..20.1513M,2001indu.book..163M,Moorhead2017,2019JSpRo..56.1531M}; if spacecraft operators are to mitigate the increased risk of damage posed by a meteor shower, they must know how long a shower lasts in addition to when it occurs. 

Within a set of meteor data, showers appear as time-limited concentrations in meteor radiant and speed. The time-limited requirement distinguishes meteor showers from sporadic sources; the latter appear as concentrations in Sun-centered ecliptic radiant and, to a lesser extent, speed that persist throughout the year. The sporadic sources exhibit seasonal variations in strength \citep{2006MNRAS.367..709C} but at no point do they become inactive. If a shower's Sun-centered ecliptic radiant lies near one of these sporadic sources, false positives for shower membership will occur at a higher rate. There may be no way to filter out these false positives individually, but the rate of false positives can be measured as shower ``activity'' far from the peak \citep{2016MNRAS.455.4329M}. This false positive rate can be used to measure a detected shower's signal-to-noise ratio \citep{2008Icar..195..317B}.

In order to study a given meteor shower, we generally must separate the members of that shower from a data set that contains both shower and sporadic meteors. This is generally accomplished by selecting the meteors that are most similar to some set of reference parameters for the shower; these parameters may be either a radiant, speed, and solar longitude, or they may be a set of orbital elements. Often, so-called dissimilarity, or $D$, parameters are used \citep{1963SCoA....7..261S,1981Icar...45..545D,1993Icar..106..603J,1999MNRAS.304..743V}. These parameters distill the difference between two orbits into a single number that is larger for more disparate orbits. As a result, similarity in one parameter can compensate for dissimilarity in another. Thus, meteors with more disparate radiants may be included near the time of the peak, when the solar longitude or ascending node is more similar to that of the shower. If measurement error in radiant and velocity is independent of observation time, this may narrow the apparent duration of the shower's activity.

In this paper, we demonstrate this shortening of the shower's activity using simple simulations of meteor showers. We construct our artificial meteor showers by assuming that their activity as a function of solar longitude can be described by a double-exponential profile and that both the Sun-centered radiant and geocentric speed of each shower remains constant over time. We also construct ``noise'' meteors, or potential false positives for shower association, that have the same distribution of Sun-centered radiant and geocentric speed but are equally likely to occur at any solar longitude. We then select shower members using two approaches. First, we apply a simple cut in radiant and geocentric velocity. Second, we apply the Drummond $D$ parameter \citep{1981Icar...45..545D} in conjunction with the cutoff values recommended by \cite{2001MNRAS.327..623G}. These methods are by no means exhaustive, but serve to demonstrate that in some cases, member extraction using $D$ parameters can artificially steepen the apparent activity profile.

\section{Methods}

This section describes our methods for generating simple simulated meteors, our assumed shower parameters, and the manner in which we have estimated a noise level.

\subsection{Generation of simulated meteors}

\begin{figure}
	\includegraphics{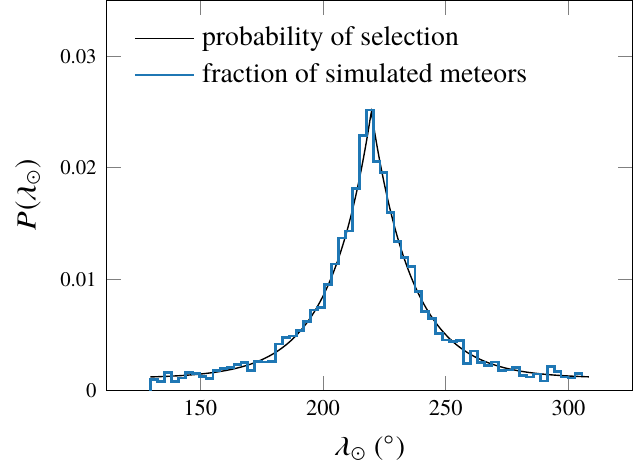}
\caption{Solar longitude probability distribution (black line) and normalized histogram of simulated solar longitude values (thick blue line) for the Southern Taurid meteor shower.}
\label{fig:fakeslon}
\end{figure}

We generate a radiant, speed, and solar longitude for each simulated meteor. First, we generate solar longitudes by assuming that the activity profile of the shower follows a double exponential profile \citep{1994A&A...287..990J,Moorhead2017}:
\begin{equation}
\mbox{ZHR} \propto 
	\begin{cases}
	10^{+B_p(\lambda_\odot - \lambda_0)} & \lambda_\odot < \lambda_0\\
	10^{-B_m(\lambda_\odot - \lambda_0)} & \lambda_\odot > \lambda_0
	\end{cases} \label{eq:shape}
\end{equation}
Thus, we adopt a form of this equation as the probability distribution for generating random solar longitude values (see Figure \ref{fig:fakeslon}).

We also assume that noise will be present in the shower at some level. We model this noise as meteors that have the same radiant and velocity distribution as shower members, but are equally likely to occur for any solar longitude. We generate these ``noise'' meteors by selecting random solar longitudes that lie within 90$^\circ$ of the peak. The noise level varies by shower and is expressed as a percentage; if, say, the assumed noise level is 20\% and the total number of simulated meteors is 5000, then 1000 will be noise meteors and 4000 will be shower meteors.

\begin{figure}
\centering
\includegraphics[width=0.45\columnwidth]{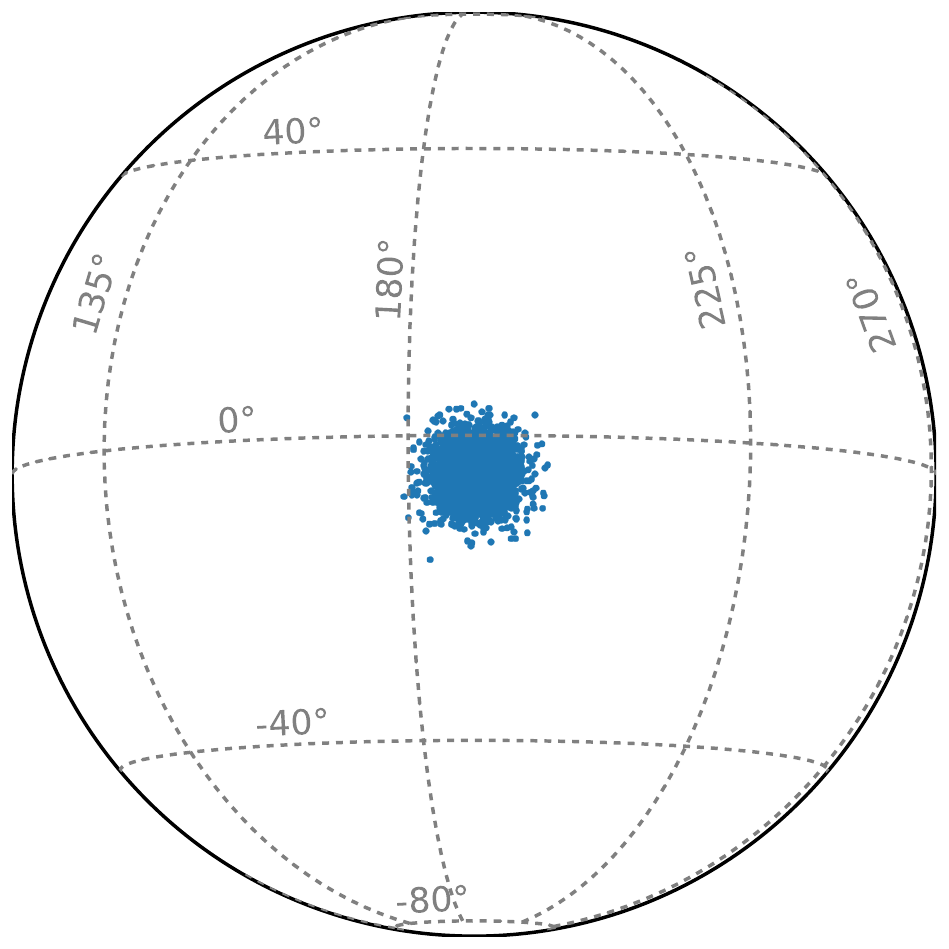}
\includegraphics[width=0.45\columnwidth]{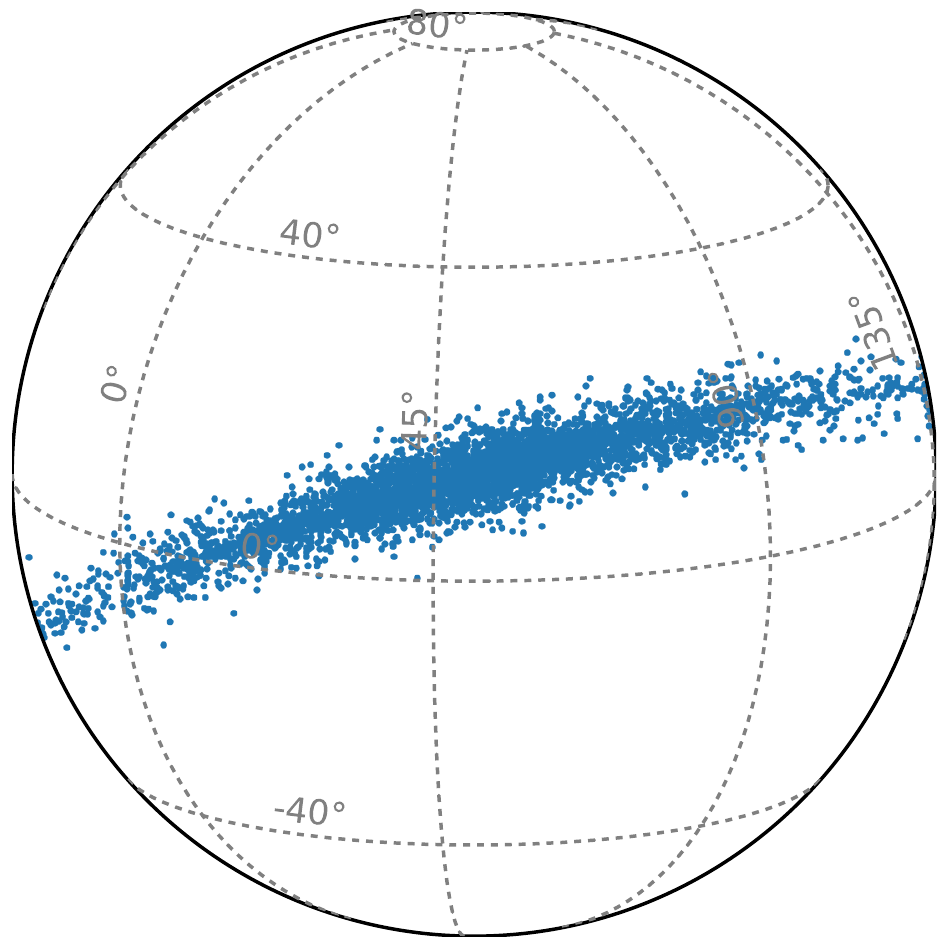}
\caption{Simulated Sun-centered ecliptic longitude and latitude (left) and right ascension and declination (right) for the Southern Taurid meteor shower.}
\label{fig:fakerad}
\end{figure}

In order to calculate simulated meteor radiants, we assume that the shower radiant remains at the same value of Sun-centered ecliptic longitude and latitude \citep[see, e.g.,][]{2010Icar..207...66B}; the typical drift of a shower radiant in these coordinates is less than one degree (Peter Brown, personal comm.). We then assume that the meteors are observed by a camera system with a relatively low level of precision, and thus that the measured radiants are scattered about the true radiant in a circularly symmetric fashion, with a normal distribution in angular radiant offset. We set the standard deviation of this offset distribution to 3$^\circ$. The distribution of Sun-centered ecliptic longitude and latitude, and the corresponding values of right ascension and declination, are depicted in Figure \ref{fig:fakerad}.

\begin{figure}
\includegraphics{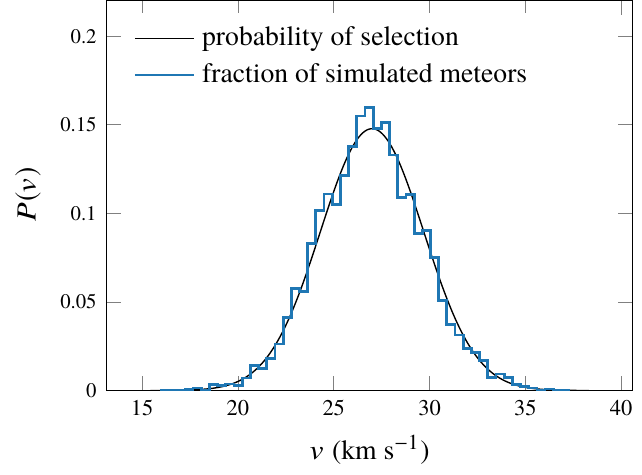}
\caption{Velocity probability distribution (black line) and normalized histogram of simulated meteoroid speeds (thick blue line) for the Southern Taurid meteor shower.}
\label{fig:fakevel}
\end{figure}

Finally, we assume that the spread in observed meteoroid speeds is also dominated by measurement error. We generate meteoroid speeds using a normal distribution centered on the true shower speed, with a standard deviation of 10\% of the shower speed (see Figure \ref{fig:fakevel}).

These simulated solar longitudes, radiants, and velocities are then converted to orbital elements by combining the meteor's geocentric velocity vector with the Earth's position at the given solar longitude in 2019 to create a complete state vector.

\subsection{Assumed shower parameters}

We will present results for two test cases: the Perseids and the Southern Taurids. The Perseids are a major meteor shower that lasts approximately a couple weeks; the Taurids are less active at any given time, but have a much longer duration. In fact, using the rise and decay time measured by \cite{1994A&A...287..990J}, Taurid activity lasts 3-6 months. Thus, we expect the use of $D$-parameters to select members of each shower to affect the Taurid activity profile more strongly than the Perseid activity profile. The full list of parameters used to simulate these showers is given in Table~\ref{tab:params}. Both the Perseids and the Southern Taurids are typically modeled as having symmetric activity profiles; that is, $B_p = B_m$ \citep{1994A&A...287..990J,2019JSpRo..56.1531M}. We therefore drop the subscripts and provide a single $B$ value for each shower. We also calculate the orbital elements for each shower at the time of the peak using the same approach described in the previous section.

\begin{table}
    \centering
    \def\arraystretch{1.3}
    \begin{tabular}{lccc}
         & & Perseids & 
            \def\arraystretch{1}
            \begin{tabular}[c]{@{}c@{}}
            Southern\\Taurids\end{tabular} \\
        \hline
        solar longitude & $\lambda_\odot$ & 
            140.05$^\circ$ & 219.7$^\circ$ \\
        &$B$ & 0.35 & 0.026 \\
        right ascension & R.A. & 
            47.2$^\circ$ & 50.1$^\circ$ \\
        declination & dec. &
            57.8$^\circ$ & 13.4$^\circ$ \\
        \def\arraystretch{0.95}%
        \begin{tabular}{@{}l@{}}
            Sun-centered\\
            ecliptic longitude\end{tabular} &
            $\lambda_\odot - \lambda_g$ &
            282.44$^\circ$ & 188.23$^\circ$ \\
        ecliptic latitude & $\beta_g$ &
            38.38$^\circ$ & -4.83$^\circ$ \\
        geocentric speed & $v_G$ & 
            59~km~s$^{-1}$ & 27~km~s$^{-1}$\\
        perihelion & $q$ & 0.96~au & 0.38~au \\
        eccentricity & $e$ & 0.93 & 0.81 \\
        inclination & $i$ & 113.3$^\circ$ & 5.3$^\circ$ \\
        argument of pericenter & $\omega$ & 151.6$^\circ$ & 113.1$^\circ$ \\
        \def\arraystretch{0.95}%
        \begin{tabular}{@{}l@{}}
            longitude of\\
            ascending node\end{tabular} & $\Omega$ & 
            140.0$^\circ$ & 39.7$^\circ$ \\
        maximum dissimilarity & $D_\mathrm{max}$ & 
            0.18 & 0.06 \\
        noise level & & 1\% & 20\%
    \end{tabular}
    \caption{Shower parameters used to generate simulated Perseid and Southern Taurid meteors. The solar longitude, right ascension, declination, and orbital elements are those of the shower at its peak or maximum activity.}
    \label{tab:params}
\end{table}

These values are assembled from a variety of sources. The peak solar longitude and $B$ values are taken from either \cite{2009JIMO...37...55S}, in the case of the Southern Taurids, or \cite{2019JSpRo..56.1531M}, in the case of the Perseids. The right ascension (R.A.), declination, and approximate geocentric speed ($v_g$) for both showers are taken from  \cite{2009JIMO...37...55S}. The cutoff $D$ parameter, which we will refer to as $D_{\mathrm{max}}$, is taken from \cite{2001MNRAS.327..623G}. These maximum $D$ values are intended to recover 70\% of the stream and vary depending on the inclination of the shower.

\subsection{Noise level}

The final parameter needed to generate our simulated data is the noise level. We roughly estimate the noise using meteor data from the NASA All-Sky Fireball Network \citep{2012pimo.conf....9C}. For each shower, we select meteors that fall within 5$^\circ$ of each shower's Sun-centered ecliptic radiant and within 20\% of the shower's geocentric velocity. We then compare the number of meteors meeting this criteria whose solar longitudes lie more than 90$^\circ$ away from the peak solar longitude to those that lie within 90$^\circ$ of the peak. We take the former to be $N_\mathrm{noise}$ and the latter to be $N_\mathrm{signal} + N_\mathrm{noise}$ for the six-month period encompassing the peak.

The Perseids produce a large number of fireballs and thus are strongly represented in our data set; they also have fairly little contamination from the sporadic background. Thus, $N_\mathrm{noise}/(N_\mathrm{signal} + N_\mathrm{noise}) \simeq 1$\% for the Perseids. In contrast, the Southern Taurids lie near the antihelion source and thus are strongly contaminated with false positives. We estimate the noise level for the Southern Taurids at about 20\%. Note that these noise levels are system dependent; systems with a lower limiting meteoroid mass, for instance, will likely detect proportionately more sporadic meteors and thus have a higher noise level.

\section{Shower extraction}

\subsection{Method 1}

We select shower members from our simulated data using one of two methods. In the first method, we select those meteors that lie within 5$^\circ$ of the Sun-centered ecliptic shower radiant and within 20\% of the geocentric shower speed. 86\% of meteors in both showers satisfy these criteria, regardless of whether they are shower meteors or ``noise'' meteors. 

\subsection{Method 2}

In the second method, we compute the orbital dissimilarity between the meteors and the shower using the Drummond $D$-parameter. This parameter is:
\begin{align}
D^2 &=  \left({\frac{q_m-q_s}{q_m+q_s}}\right)^2 + \left({\frac{e_m-e_s}{e_m+e_s}}\right)^2 +\left({\frac{I}{180^\circ}}\right)^2 + \left({\frac{e_m+e_s}{2} \cdot \frac{\theta}{180^\circ}}\right)^2 \, .
\end{align}
where $q$ is perihelion distance, $e$ is orbital eccentricity, and $i$ is orbital inclination. The subscripts indicate whether the parameter is that of the shower ($s$) or of an individual meteor ($m$). The terms $I$ and $\theta$ give the angle between the two orbital planes and the angle between the lines of apsides, respectively:
\begin{align}
I &= \arccos{\left[{\cos{i_m} \cos{i_s} + \sin{i_m} \sin{i_s} \cos(\Omega_m - \Omega_s)}\right]} \\
\theta &= \arccos [\sin \eta_m \sin \eta_s + \cos \eta_m \cos \eta_s \cos(\gamma_s - \gamma_m)] \, .
\end{align}

The symbols $\gamma$ and $\eta$ give the ecliptic longitude and latitude of each orbit's perihelion, which can be computed as follows:
\begin{align}
\gamma &= \Omega + \arctan (\cos i \tan \omega) \\
\eta &= \arcsin (\sin i \sin \omega) \, .
\end{align}
where $\omega$ is the argument of pericenter and $\Omega$ is the longitude of the ascending node.

For membership, we require $D < 0.18$ for the Perseids and $D < 0.06$ for the Southern Taurids. Although these cutoff values correspond to 70\% shower retrieval according to \cite{2001MNRAS.327..623G}, we do not achieve 70\% recovery. Instead, this method recovers 52\% of Perseids and 25\% of Taurids (not including noise meteors). This may be because we have simulated measurement errors larger than those encountered by \cite{2001MNRAS.327..623G} in the AMOR (Advanced Meteor Orbit Radar) data. In general, the shower retrieval thresholds will vary between networks and even between showers; those wishing to optimize their shower member retrieval rate should characterize its dependence on the $D$ cutoff value within their data set \citep{2016MNRAS.455.4329M}.

\subsection{Results}

\begin{figure}
\includegraphics{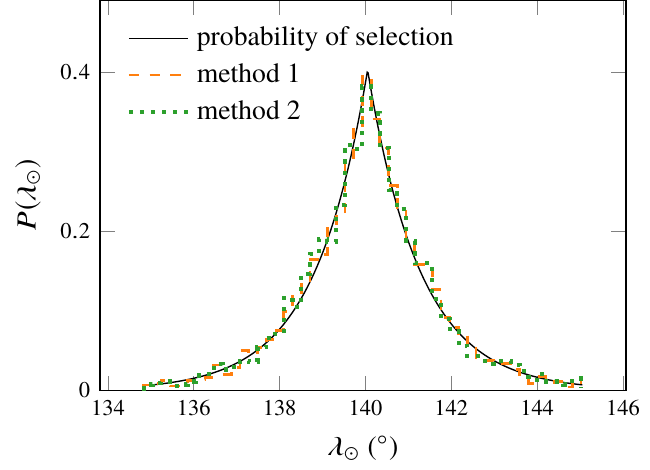}
\caption{Solar longitude probability distribution (black line) and normalized histogram of selected simulated solar longitude values for the Perseid meteor shower. Two selection methods are shown: method 1 (dashed orange line) selects meteors based on their radiant and velocity, and method 2 (dotted green line) uses an orbital dissimilarity parameter.}
\label{fig:perslon}
\end{figure}

\begin{figure}
\includegraphics{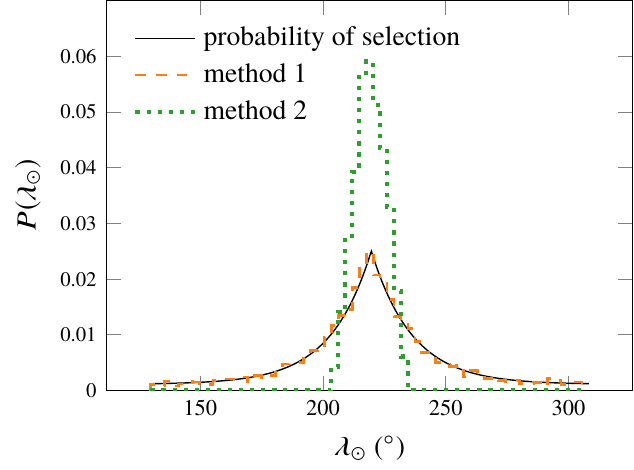}
\caption{Solar longitude probability distribution (black line) and normalized histogram of selected simulated solar longitude values for the Southern Taurid meteor shower. Two selection methods are shown: method 1 (dashed orange line) selects meteors based on their radiant and velocity, and method 2 (dotted green line) uses an orbital dissimilarity parameter.}
\label{fig:staslon}
\end{figure}

Figures \ref{fig:perslon} and \ref{fig:staslon} show the distribution of simulated meteor solar longitudes selected using these two methods. In the case of the Perseids, both methods preserve the activity profile of the shower. However, in the case of the Southern Taurids, the use of an orbital dissimilarity parameter (method 2) significantly alters the shape of the activity profile. In fact, if we attempt to measure $B$ from the distribution labeled ``method 2'' in Figure \ref{fig:staslon}, we obtain a value of about 0.08, which is between 3 and 4 times the value used to generate the simulated data.

\section{Conclusion}

We have demonstrated that the use of orbital dissimilarity parameters to select members of a meteor shower can result a shower activity profile that is artificially brief and steep in shape. This effect is not significant for short-to-medium duration showers such as the Perseids, but can be quite significant for long-duration showers such as the Southern Taurids. It will occur for any method that incorporates time, solar longitude, or longitude of the ascending node into a single measure of shower member likelihood. This includes not only the Drummond $D$, but most other variants \citep[including][]{1963SCoA....7..261S,1993Icar..106..603J,1999MNRAS.304..743V,2002dnac.conf..365N,2008EM&P..102...73J}.

It may be beneficial to preferentially omit meteors that occur far from the peak; for instance, the use of the Drummond $D$ parameter to select Southern Taurids from our simulated date retrieves 25\% of simulated Taurids, but only 2\% of the simulated noise. Thus, a $D$-parameter-based selection method can reduce the quantity of sporadic interlopers and perhaps improve the calculation of an average orbit. It is therefore important to tailor one's shower selection method to one's end goal.

One can sidestep the bias discussed here in one of two ways. The first solution is to disregard time, solar longitude, or longitude of ascending node when assessing shower membership. For instance, the wavelet coefficient presented by \cite{2008Icar..195..317B} uses only (Sun-centered ecliptic) radiant and velocity to detect showers. Their search is conducted once per degree of solar longitude and similar wavelet peaks on adjacent days are linked afterwards. As a result, temporal clustering does not enter into the initial selection method and will not affect the activity profile. 

A second, equivalent approach is to use a set of reference shower orbits (or radiants and velocities) instead of a single, static orbit. If the solar longitude or longitude of ascending node of the reference orbits varies over time in a manner that accurately traces the shower's behavior, and similarity to any of these reference orbits qualifies the meteor as a shower member, the temporal bias we've discussed will be avoided. An example of this approach can be found in \cite{2018P&SS..154...21J}, who generated a ``look-up table'' that describes how shower elements drift over time.

\bibliography{local}

\end{document}